\documentclass[aps,prb,preprint,preprintnumbers,amsmath,amssymb]{revtex4} 
\usepackage{graphicx}
\usepackage{bm}
\usepackage[final]{pdfpages}

\begin{document}
\title{Anomalous Cooper pair interference on Bi$_2$Te$_3$ surface}
\date{\today}

\author{Jie Shen}\thanks{These authors contribute equally to this work.}
\author{Yue Ding}\thanks{These authors contribute equally to this work.}
\author{Yuan Pang}
\author{Fan Yang}
\author{Fanming Qu}
\author{Zhongqing Ji}
\author{Xiunian Jing}
\author{Jie Fan}
\author{Guangtong Liu}
\author{Changli Yang}
\author{Genghua Chen}
\author{Li Lu} \email[Corresponding authors: ]{lilu@iphy.ac.cn} \affiliation{Daniel Chee Tsui Laboratory, Beijing National Laboratory for Condensed Matter Physics, Institute of Physics, Chinese Academy of Sciences, Beijing 100190, People's Republic of China}


\begin{abstract}
\textbf{It is believed that the edges of a chiral $p$-wave superconductor host Majorana modes, relating to a mysterious type of fermions predicted seven decades ago \cite{Majorana, Majorana_return}. Much attention has been paid to search for $p$-wave superconductivity in solid-state systems \cite{organic_SC, UPt3, SrRuO_Liu_Ying_SrRuO_Japan, Willett}, including recently those with strong spin-orbit coupling (SOC) \cite{Kouwenhoven2006, Ando, Yang_Fan, Kowenhoven, Jia, Xu}. However, smoking-gun experiments are still awaited. In this work, we have performed phase-sensitive measurements on particularly designed superconducting quantum interference devices constructing on the surface of topological insulators Bi$_2$Te$_3$, in such a way that a substantial portion of the interference loop is built on the proximity-effect-induced superconducting surface. Two types of Cooper interference patterns have been recognized at low temperatures. One is $s$-wave like and is contributed by a zero-phase loop inhabited in the bulk of Bi$_2$Te$_3$. The other, being identified to relate to the surface states, is anomalous for that there is a phase shift between the positive and negative bias current directions. The results support that the Cooper pairs on the surface of Bi$_2$Te$_3$ have a 2$\pi$ Berry phase which makes the superconductivity $p_x+ip_y$-wave-like. Mesoscopic hybrid rings as constructed in this experiment are presumably arbitrary-phase loops good for studying topological quantum phenomena.}
\end{abstract}

\maketitle

Topological insulators (TIs) \cite{123} are a class of insulators with topologically protected conducting surface due to strong SOC. The SOC interlocks the momentum and spin degrees of freedom of the electrons in the surface states, so that the motion of the electrons there appears to be helical, acquiring a Berry phase. The helical surface states survive in the presence of a conducting bulk, as revealed by both angular resolved photon emission spectroscopy \cite{XJZhou_ARPES_ZXshen} and electron transport measurements \cite{nethelands_ZhangCi}.

When the helical electrons pair up via, e.g., superconducting proximity effect, Fu and Kane proposed that they may behave like a spinless $p_x+ip_y$-wave superconductor \cite{Fu_liang}. Figure 1\textbf{a} illustrates how the pair correlation is delivered from an $s$-wave superconductor (colored in dark blue) to the surface and the bulk of a three-dimensional (3D) TI candidate (i.e., with bulk states) via proximity effect. While the induced superconductivity in TI in the vicinity of the $s$-wave superconductor keeps more or less conventional $s$-wave-like (colored in light blue in Fig. 1\textbf{a}), at a farther distance on the TI surface, however, only those Cooper pairs formed of helical electrons survive (colored in orange in Fig. 1\textbf{a}).

In general, the wavefunction of a Cooper pair should include the information of the external mass-center orbital part, the internal relative orbital part and the spin part of two paired electrons. This keeps to be true in the presence of SOC. Inherited from each helical electron who encounters a Berry phase of $\pi$ in the circulation mode along a mesoscopic ring \cite{QuPRL}, winding the mass center of a Cooper pair along the ring rotates the two spins of paired helical electrons in the same angular direction, doubling the total Berry phase accumulated regardless that they are of singlet pairing (further explanations can be found in the supplementary material). A Berry phase of 2$\pi$ over a 360$^{\circ}$-turn on a 2D surface gives rise the $p_x+ip_y$-wave-like character to the transport properties of the Cooper pairs, regardless that both the total spin angular momentum and the relative orbital angular momentum are zero.

\begin{figure}
\includegraphics[width=0.75 \linewidth]{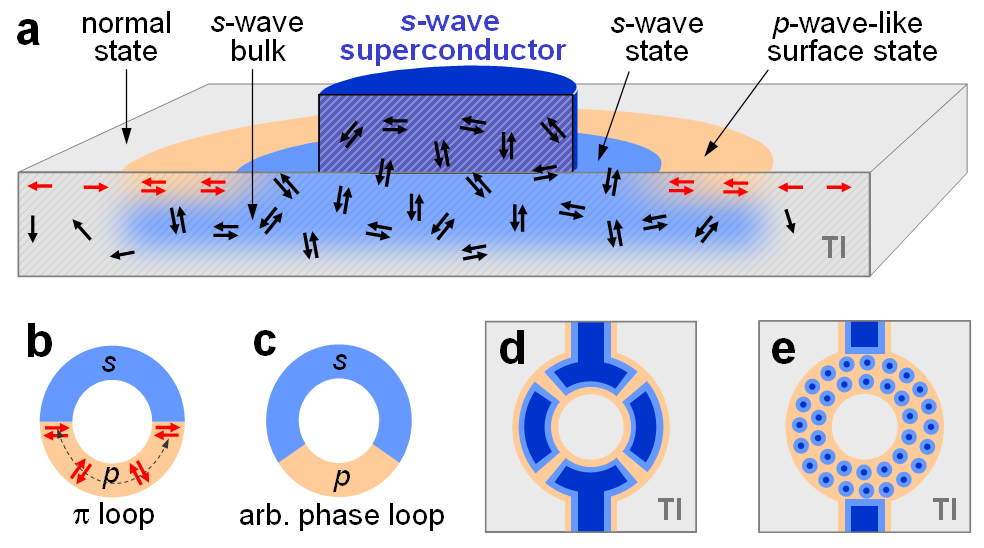}
\caption{\label{fig:fig1} {(color online) \textbf{a}, A cross-section view on the development of proximity-effect-induced superconductivity from an $s$-wave superconductor (the dark blue part) to a topological insulator (TI) with surface and bulk states. The arrows represent the spins of paired and unpaired electrons. While the conventional $s$-wave character is kept within a certain distance from the interface (the regions colored in light blue), those paired helical electrons (the red arrows) at a farther distance on the surface form a $p_x+ip_y$-like spin-singlet superconducting state (colored in orange), due to picking up the 2$\pi$ Berry phase generated by the spin-orbit coupling. \textbf{b}, Illustration of a $\pi$-loop \cite{XLQi}, where a Berry phase of $\pi$ is contributed by the half ring of $p_x+ip_y$-wave superconductor, and is encountered in the circulation mode of the Cooper pairs along the ring. \textbf{c}, Illustration of an arbitrary-phase loop, where the phase shift encountered by the Cooper pairs depends on the turning angle of the mode in the $p_x+ip_y$-wave superconductor. \textbf{d} and \textbf{e}, Two experimentally accessible variations of arbitrary-phase loops utilizing the superconducting proximity effect at the surface of a 3D TI, in which the $s$ and $p_x+ip_y$-like segments are cut to small pieces (multiple segments and small grains, respectively) and mixed up together.}}
\end{figure}

To determine the pairing symmetry of the induced superconductivity in TIs, one needs to perform phase-sensitive experiments similar to those determined the $d$-wave pairing symmetry in cuprates \cite{Harlingen_CCTsuei}. Experimentally, superconducting proximity effect has been observed between conventional $s$-wave superconductors such as Al, Nb, Sn, Pb and 3D TIs such as Bi$_2$Se$_3$ and Bi$_2$Te$_3$. However, the Fraunhofer diffraction patterns of single Josephson junctions and the interference patterns of superconducting quantum interference devices (SQUIDs) were mostly $s$-wave like, i.e., the critical supercurrents maximizes at zero magnetic flux \cite{QuFM, Goldhaber-Gordon, Brinkman, yangfan1}. One exception was observed on a large-area SQUID where a tilted interference pattern was caused by the magnetic flux generated by the bias current \cite{Brinkman_APL}.

The absence of $p_x+ip_y$-like signature in above mentioned experiments is actually expected, since the modes of Cooper pairs pass through the induced superconducting regions (which serve as the junctions) straightly, picking up no Berry phase. In order to probe the Berry phase, one needs to form an interference loop incorporating curved segments of $p_x+ip_y$-like superconductors (Fig. 1 \textbf{b} \cite{XLQi} and \textbf{c}). The use of discrete or granular $s$-wave superconductors (Fig. 1 \textbf{d} and \textbf{e}) makes it possible to bridge the interference loop on one hand, and to allow the phase of the Cooper pair to vary/accumulate in the $p_x+ip_y$-like regions along the loop on the other hand.

\begin{figure}
\includegraphics[width=0.8 \linewidth]{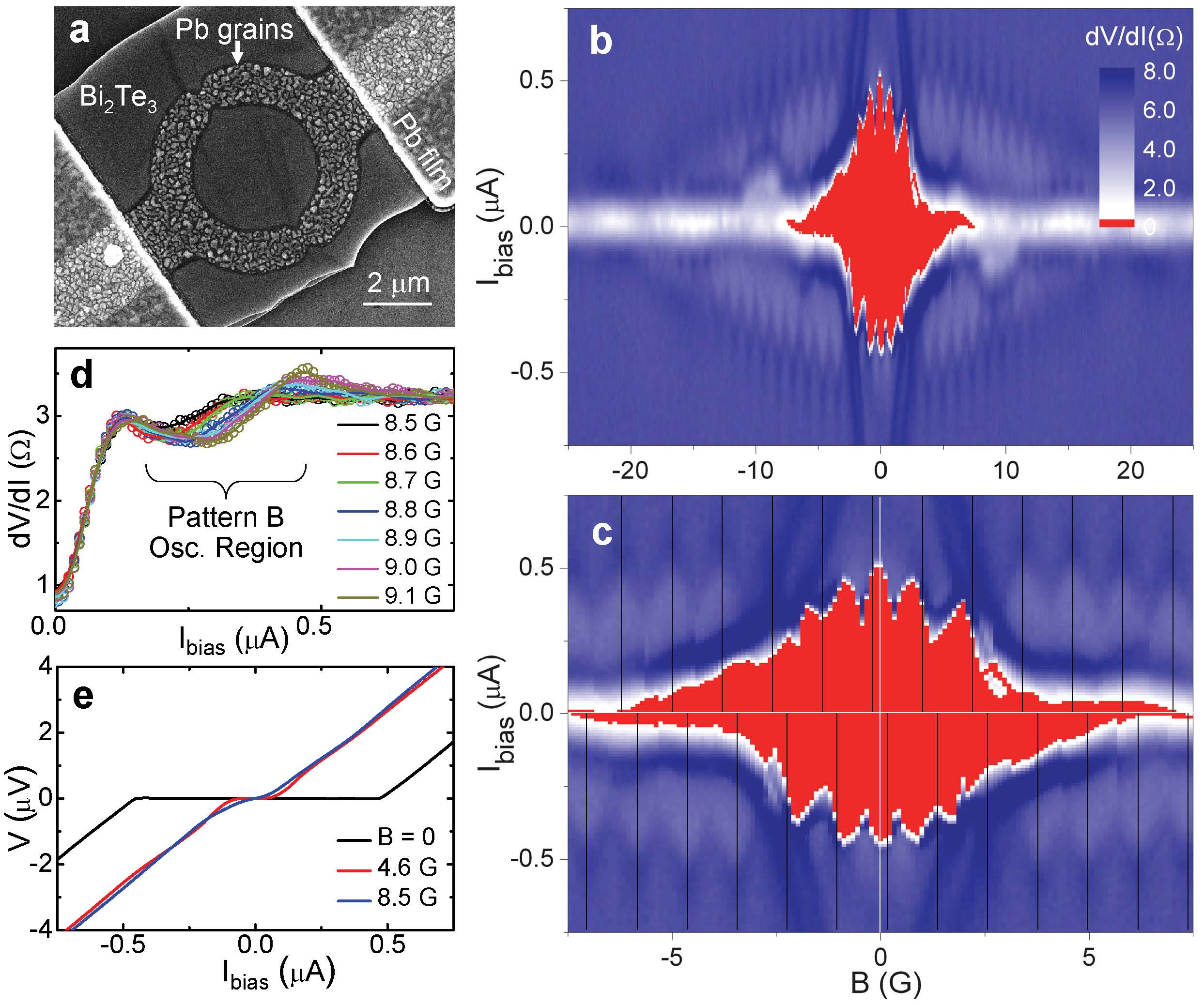}
\caption{\label{fig:fig2} {(color online) Cooper pair interference in a superconducting quantum interference device composed of Pb grains on the surface of a Bi$_2$Te$_3$ single crystalline flake. \textbf{a}, A scanning electron microscope image of the device. \textbf{b}, Differential resistance $dV/dI$ of the device measured at 30 mK, as functions of bias current $I_{\rm bias}$ and magnetic field $B$ perpendicular to the device plane. \textbf{c}, The central part of \textbf{b}. Two oscillation patterns can be recognized. The vertical lines are guide to the eyes showing that the peak positions of one of the patterns (pattern B) are horizontally shifted by an amount of $2\delta\approx 0.37\times 2\pi$ between its positive and negative bias current directions. \textbf{d} and \textbf{e}, $dV/dI-I_{\rm bias}$ and $V-I_{\rm bias}$ curves, respectively, taken at several different fields at 30 mK.}}
\end{figure}

We have fabricated Pb-grain SQUIDs on the surface of Bi$_2$Te$_3$ flakes (i.e., the design in Fig. 1\textbf{e}) in several different Pb coverage ratios, and studied the interference of Cooper pairs via electron transport measurements down to 30 mK in a dilution refrigerator. We have also explored the multi-segment design shown in Fig. 1\textbf{d}. The results can be found in the supplementary material.

Figure 2\textbf{a} shows the scanning electron microscope image of a Pb-grain SQUID. Between the two bulky Pb electrodes there is a ring composed of Pb grains of 10-50 nm in diameter and 10-20 nm in height. These grains were formed naturally when Pb was sputtered onto the Bi$_2$Te$_3$ surface. Figures 2\textbf{b} and 2\textbf{c} show the differential resistance of the device measured at 30 mK. Two interference patterns can be recognized. One represents a zero-resistance state ($dV/dI\leq 0.2\Omega$, marked in red), hereafter called pattern A. The other inhabits on a low but finite-resistance state, with a whitish-blue color, hereafter referred to as pattern B. The periods of the patterns, $\Delta B_A=1.0\pm 0.1$ G for pattern A (best seen in Fig. 3\textbf{h}) and $\Delta B_B=1.15\pm 0.1$ G for pattern B, correspond to two slightly different areas of $S_A=21\pm 2 \mu$m$^2$ and $S_B=18.1\pm 1.5 \mu$m$^2$, respectively, if estimated using $\phi_0\equiv h/2e=S\Delta B$, where $\phi_0$ is the flux quanta. The oscillations in pattern B are relatively regular, with a period which is in good agreement with the effective area of the ring \cite{Josephson}: $S=\pi D_{in}D_{out}/4\approx 18.8 \mu$m$^2$, where $D_{in}=4 \mu$m and $D_{out}=6 \mu$m are the inner and outer diameters of the ring respectively.

The Pb coverage ratio of the device shown in Fig. 2 is 71$\pm$2\%. By controlling the deposition time, devices with different Pb coverage ratios were also obtained and studied. It appears that the devices with higher Pb coverage ratios (e.g., the ones with continuous Pb films \cite{QuFM} or with a Pb network, see Fig. S2 of the supplementary material) show only pattern A, whereas the devices with Pb grains of lower coverage ratios mostly often show only pattern B. The results are summarized in Table S1 of the supplementary material.

Pattern A should arise from a conventional 0-loop, since its highest peak is located at zero magnetic field in both positive and negative bias current directions. On the contrary, pattern B has the following anomalous features. (i) There is a phase shift between the positive and negative current halves. (ii) It inhabits on a finite resistance state. (iii) It appears to be more robust than pattern A, i.e, the oscillations are more uniform, with a wider envelop against magnetic field and survivable to higher temperatures. Also, as we will show later, pattern B survives in larger parallel magnetic fields. Adding to the robustness, pattern B was observed on all the seven working Pb-grain devices out of seven that were investigated, whereas pattern A was observed to coexist only in two of the seven.

\begin{figure}
\includegraphics[width=0.6 \linewidth]{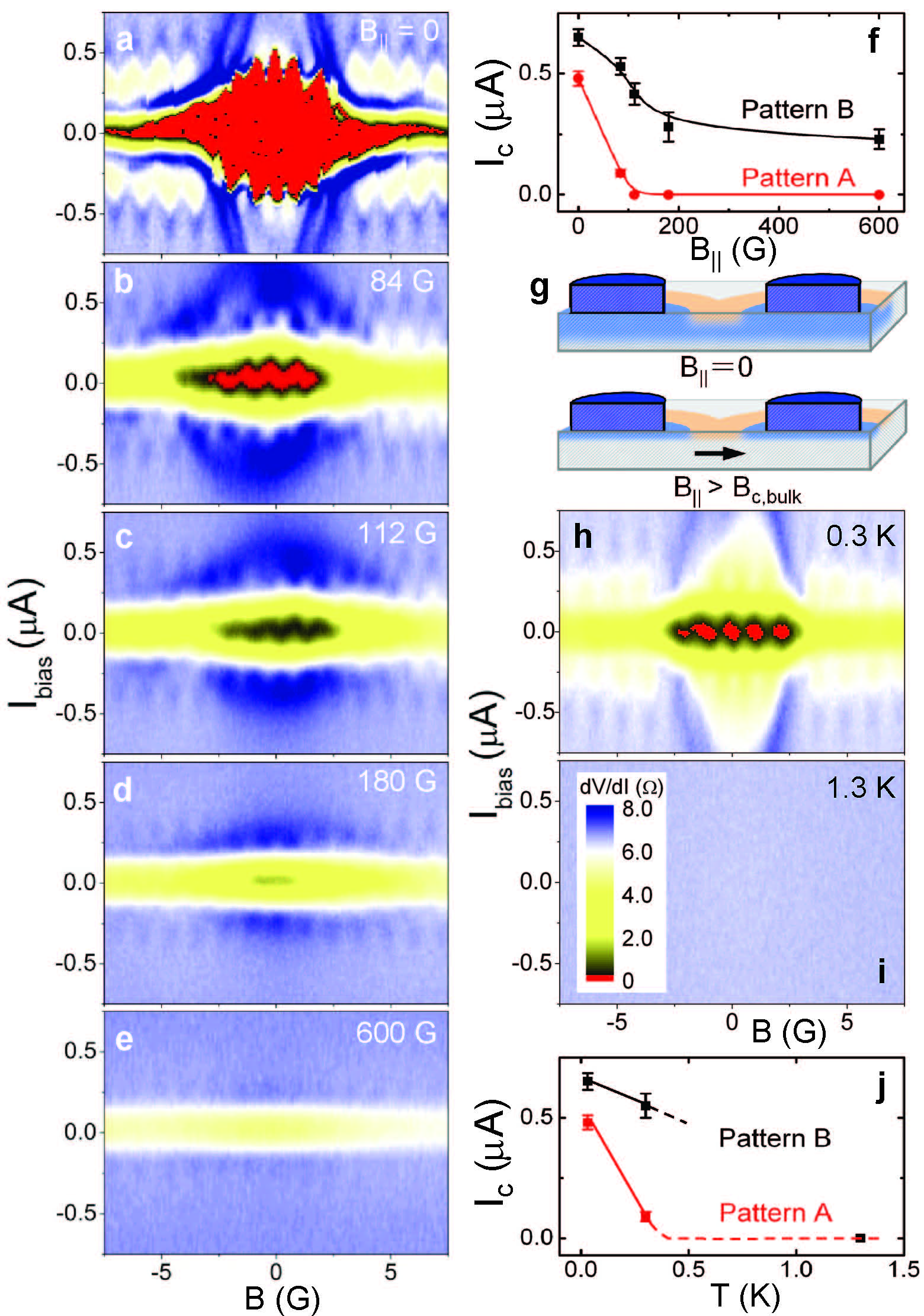}
\caption{\label{fig:fig3} {(color online) \textbf{a} - \textbf{e}, Evolution of the patterns shown in Fig. 2 as a function of parallel magnetic fields $B_{\|}$, measured at 30 mK. The data are plotted according to the color scale in Fig. 3\textbf{i}. Pattern A is suppressed when $B_{\|}\geq 100$ G. Pattern B remains to be visible at $B_{\|}\simeq 600$ G. \textbf{f}, The amplitudes of the two patterns as a function of parallel magnetic field at 30 mK. \textbf{g}, A cartoon illustrating that the bulk superconductivity (colored in light blue) is easier to be suppressed than the surface superconductivity (colored in orange) in a parallel magnetic field. \textbf{h} - \textbf{j}, Evolution of the two patterns with varying temperature at $B_{\|}=0$.}}
\end{figure}

Based on the fact that the surface electrons and the bulk electrons manifest distinctively in their Shubnikov-de Haas oscillations \cite{nethelands_ZhangCi}, we speculate that the induced superconducting states at the surface and in the bulk, as well as the interplay between them, give rise to the two distinctive interference patterns. In order to confirm the surface and bulk contributions to the two observed patterns, we applied a parallel magnetic field trying to suppress the superconductivity in the bulk of Bi$_2$Te$_3$ flake.

Figures 3\textbf{a} to 3\textbf{e} show the evolution of the patterns in several parallel magnetic fields $B_{\|}$. While pattern A is suppressed at $B_{\|}\geq 100$ G, pattern B still survives when $B_{\|}\simeq 600$ G. The field dependencies of their oscillation amplitudes are summarized in Fig. 3\textbf{f}. In Figs. 4\textbf{b} to 4\textbf{f} we further show the data observed on another device, whose pattern B survives in a parallel magnetic field of a few thousands Gauss at which the induced bulk superconductivity in TI should have been suppressed completely. Figure 3\textbf{g} is a cartoon illustrating that the superconductivity in the bulk (colored in light blue) is easier to be suppressed than that in the surface shell (colored in orange) in a parallel magnetic field, for the reason explained below.

Because less diamagnetic energy is needed in field penetration, the effective critical field of a thin superconducting plate in parallel magnetic fields can be greatly enhanced \cite{Tinkham}: $B^*_{\rm c}=\sqrt{24}B_{\rm c}\lambda_{\rm L}/d$, where $B_{\rm c}$ and $\lambda_{\rm L}$ are the critical field and the London penetration depth of the material in its bulk form, respectively, and $d$ is the thickness of the thin plate. For the induced superconducting surface and bulk in Bi$_2$Te$_3$, therefore, the ratio between their effective critical fields is $B^*_{\rm c,surf}/B^*_{\rm c,bulk}\simeq d_{\rm bulk}/d_{\rm surf}$ if neglecting any helicity-related difference. It roughly agrees with the observations in this experiment, i.e., $B^*_{\rm c,surf}$ being several thousands Gauss, $B^*_{\rm c,bulk}\simeq$100 G, $d_{\rm bulk}\simeq 100$ nm, and $d_{\rm surf}$ being the thickness of the surface state, which is a few nanometers \cite{surface_thickness_meas}. We thus believe that pattern B is contributed by an interference loop incorporating the superconducting surface, when the superconducting bulk beneath the Pb islands subsides and disconnects from each other.

\begin{figure}
\includegraphics[width=0.55 \linewidth]{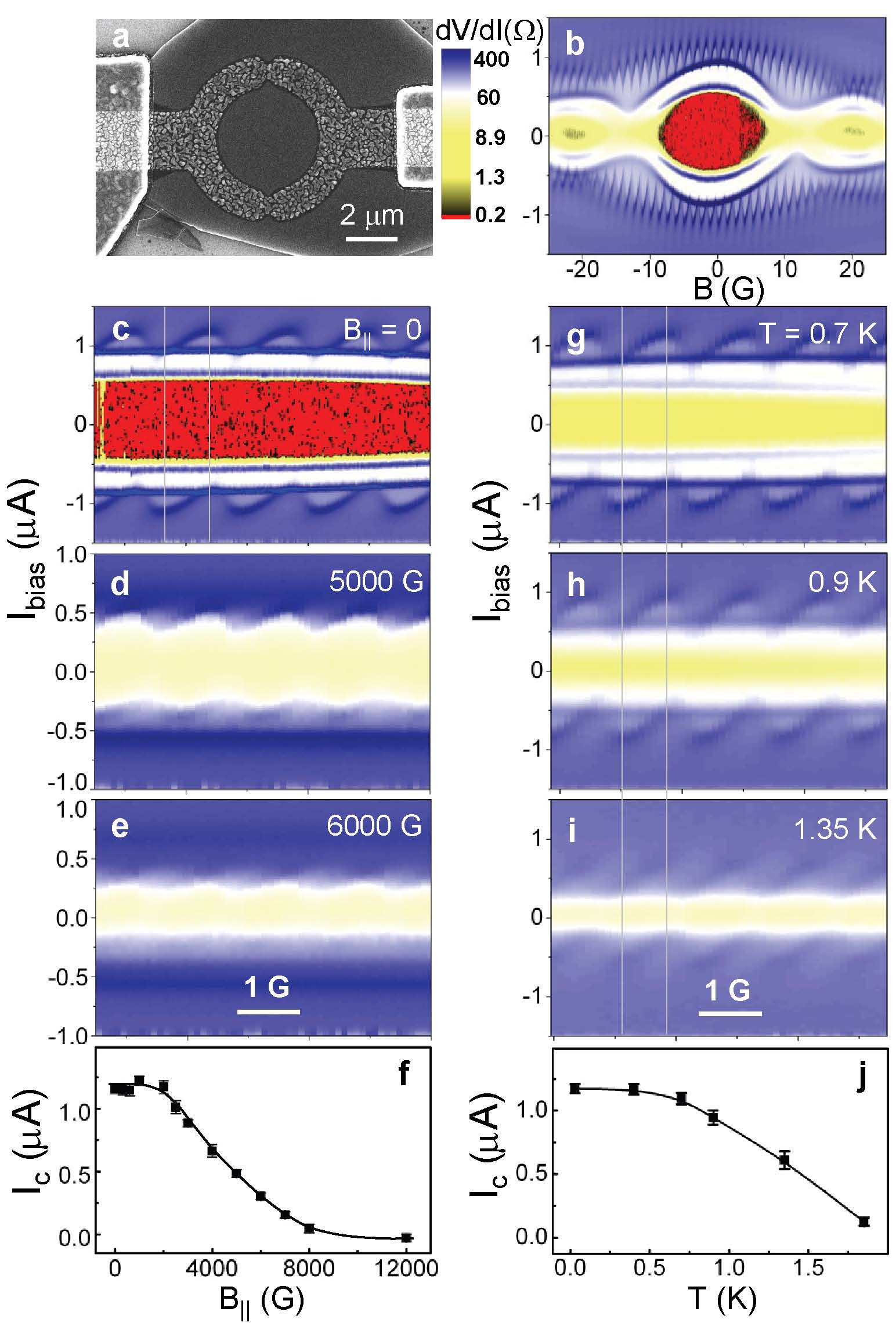}
\caption{\label{fig:fig4} {(color online) Cooper pair interference in a second Pb-grain SQUID. \textbf{a} A scanning electron microscope image of the device. \textbf{b} Differential resistance of the device measured at 30 mK. Its zero-resistance state (with $dV/dI\leq 0.2\Omega$, colored in red) shows only a Fraunhofer diffraction pattern but no interference oscillations. Nevertheless, pattern B is still there, demonstrating a phase shift of $\delta\approx 0.69\pi$. \textbf{c} - \textbf{f}, Evolution of pattern B in parallel magnetic fields measured at 30 mK (only those data within a window of a few Gauss near $B=0$ in \textbf{b} are traced). \textbf{g} - \textbf{j}, Evolution of pattern B with varying temperature at $B_{\|}=0$. The vertical lines are guide to the eyes showing that the phase shift remains almost constant.}}
\end{figure}

In the following, we will discuss whether the observed shift of pattern B reflects an intrinsic character of the superconducting surface, or rather caused by inhomogeneity and the random formation of multiple ($\geq$3) junctions unevenly distributed on the ring (i.e., mimic a Josephson junction ratchet \cite{Ratchet}).

The fact that pattern B is more robust than pattern A indicates that the induced superconducting surface is much easier to form an interference loop than the induced superconducting bulk, though their superconductivity stem from the same Pb grains. It indicates that the Cooper pairs have a longer coherence length on the surface than in the bulk, in agreement with the results of existing experiments \cite{Brinkman}. Our previous studies also show that the superconductivity spreads from Pb to a distance of several micrometers into Bi$_2$Te$_3$ \cite{QuFM, yangfan1}, a length which is long enough to bridge the Pb grains in this experiment to form a continuous superconducting loop. Therefore, for those Pb-grain devices whose bulk has already formed a superconducting loop and demonstrated an unshifted interference pattern, it is reasonable to believe that their surface has formed an even more uniform superconducting loop.

On the contrary, if pattern B had not been caused by well-defined junctions but the multiple junctions formed of inhomogeneity origin, the peak positions, which depend on the fine balance between the critical supercurrents of different junctions, would have varied a lot with temperature. Even the number of junction would have changed with temperature. This is not true from our data. In Fig. S5 of the supplementary material, we further show the data obtained on a geometrically well-defined symmetric four-segment SQUID (following the design in Fig. 1\textbf{d}). In spite of having a conventional pattern A, the device also exhibited an anomalous pattern B. Based on all these facts, we conclude that pattern B is an intrinsic property of the superconducting surface, not caused by extrinsic mechanisms such as a rachet.

At first glance, pattern B in Fig. 4 and Fig. S5 of the supplementary material appears to be tilted. However, it is not true. A tilted interference pattern usually arises when the applied current exerts a significant amount of magnetic flux to the SQUID loop. For our devices, the flux generated by the applied current of the order $I_{\rm c}=1\mu$A cannot exceed $LI_{\rm c}/2\thickapprox 0.0007\phi_0$ (where $L\approx 3$ pH is the inductance of the ring with a radius of 2.5 $\mu$m), so that current-induced tilting is negligible. Instead of being tilted, all the peaks in pattern B are actually shifted horizontally by a same amount regardless of their height, which can be best seen from the data of the four-segment SQUID shown in the supplementary material (Fig. S6).

A shift in peak position implies that there is an additional phase added to the phase quantization condition of the loop \cite{Josephson}: $\phi_1+\phi_2+2\pi{\it\Phi}/\phi_0\pm\delta=2\pi N$, where $\phi_1$ and $\phi_2$ are the phase differences across the two weak links on the arms, ${\it\Phi}$ is the applied magnetic flux, $N$ is an integer, and $\pm\delta$ is presumably the Berry phase acquired by the Cooper pairs from the $p_x+ip_y$-like segments of the loop in the clockwise/counterclockwise (cw/ccw) circulation modes.

In a ring with mixed $s$-wave and $p_x+ip_y$-wave-like segments, the acquired Berry phase is in general not necessary $\pi$ (Fig. 1\textbf{b}), but an arbitrary value depending on the accumulated turning angle of the mode in the $p_x+ip_y$-wave-like region (See Fig. 1\textbf{c}. More discussions on arbitrary-phase loops and fractional modes can be found in the supplementary material). Adding an arbitrary phase to the phase quantization condition would shift the free energy minimum of the device away from the zero flux point, forming a symmetric quantum double-well for its cw/ccw circulation modes. The device thus undergoes spontaneous symmetry breaking, associated with a chiral edge supercurrent. In the presence of inter-well tunneling, however, the system further becomes a two-level system, so that the time-reversal symmetry is restored. Applying a bias current tilts the double-well potential via a given asymmetric current distribution of the device, changes the population/dwelling time of the quantum state in the two wells as illustrated in Fig. S7 of the supplementary material.

Another unusual feature of pattern B is that it inhabits on a finite resistance state, superimposed on the bulk contribution. A finite resistance state at low temperatures is often caused by phase fluctuations and dissipations, especially in superconductors with reduced dimensions. In Josephson junctions with a critical supercurrent of 1 $\mu$A or smaller, like the ones in this experiment, phase diffusion often happens via macroscopic quantum tunneling (MQT) \cite{Josephson}. According to Fig. 2 \textbf{d} and \textbf{e}, the voltage across the junctions in pattern B region is
$1-2$ $\mu$V, which is comparable to the thermal energy of $\sim$30 mK in the measurements. It indicates that phase diffusion happens probably via thermally-assisted MQT. From the Josephson equation $\langle\dot{\phi}\rangle=2eV/h$, the phase changing rate is $0.5-1$ GHz, or the dwelling time of the phase particle in the washboard potential is around $1-2$ nS, which is long enough to establish Cooper pair interference along the loop. In addition to phase diffusion, Little-Parks mechanism \cite{Little-Parks} might also be responsible for the finite-resistance state. Further study is needed to clarify this issue.

To conclude, anomalous Cooper pair interference has been observed on SQUIDs whose a substantial portion of interference loop is built on the superconducting surface of Bi$_2$Te$_3$. The results suggest that the Cooper pairs there on the surface have a Berry phase. Mesoscopic hybrid rings constructed in this experiment are likely arbitrary phase loops in general, and would serve as controllable model systems with improved preparation procedures, good for studying exotic physics of topological quantum states such as exploring for Majorana fermions and topological quantum computation.

\textbf{The method}: The Bi$_2$Te$_3$ flakes of $\sim$100 nm thick were exfoliated from high-quality single crystals grown by Bridgman method, and were then transferred onto Si/SiO$_2$ substrates. Bulk Pb films of $\sim$200 nm thick and grain-like Pb structure were fabricated onto the flakes via standard e-beam lithography, magnetron sputtering and lift-off techniques. The electron transport measurements were performed at low temperatures down to 30 mK in a dilution refrigerator. The differential resistance of the devices was measured as functions of both dc bias current and applied magnetic fields in both the perpendicular and parallel directions.

\begin{acknowledgments}
We would like to thank T. Xiang, L. Fu, Y. Liu, Z. Fang, X. Dai, Y. P. Wang, S. Y. Han, S. P. Zhao, G. M. Zhang, X. C. Xie, R. Du, Q. Niu, S. C. Zhang, L. Yu and X. L. Qi for stimulative discussions. This work was supported by the National Basic Research Program of China from the MOST under the Contracts No. 2009CB929101 and No. 2011CB921702, by the NSFC under Contracts No. 91221203, No. 11174340 and No. 11174357, and by the Knowledge Innovation Project and the Instrument Developing Project of CAS.
\end{acknowledgments}

\vspace {1 cm}

{\bf Author contributions}

\vspace {0.5 cm}

L.L. conceived and designed the experiments. F.M.Q. grew the crystals. Y.D. and J.S. fabricated the devices. J.S., Y.D. and Y.P. performed the measurements. Z.J., J.F. and X.N.J. helped on the experiment. G.T.L. and C.L.Y. participated in the discussions. J.S. and L.L. prepared the manuscript.

\vspace {1 cm}

{\bf Additional information}

\vspace {0.5 cm}

The authors declare no competing financial interests. Correspondence should be addressed to L.L.



\section*{Supplementary material}

# Supplementary Materials for
# "Anomalous Cooper pair interference on $Bi_2Te_3$ surface"

Jie Shen*, Yue Ding*, Yuan Pang, Fan Yang, Fanming Qu, Zhongqing Ji, Xiunian Jing, Guangtong Liu, Jie Fan, Changli Yang, Genghua Chen and Li Lu†

*Daniel Chee Tsui Laboratory, Beijing National Laboratory for Condensed Matter Physics & Institute of Physics, Chinese Academy of Sciences, Beijing 100190, People's Republic of China*

\* These authors contributed equally to this work.
† Corresponding authors: lilu@iphy.ac.cn

Contents:

1. Material and devices characterizations
2. More devices with different Pb coverage ratios
3. The upper critical fields of Pb thin films and Pb grains
4. Interplay and competition between pattern A and pattern B
5. Investigation on a four-segment SQUID
6. Is pattern B shifted or tilted?
7. Spontaneous symmetry breaking and current-driven symmetry breaking in a double-well system
8. Further explanations on the Berry phase and the possible consequences of having arbitrary phase loops in mesoscopic hybrid rings

1. Material and devices characterizations

$Bi_2Te_3$ single crystals used in this experiment were grown by Bridgman method and were proven to be of high quality by X-ray diffraction and electron transport measurements. The mobility of the carriers is 2000 - 5000 $cm^2$/Vs. Relevant data can be found in Ref. 1.

After the crystals were mechanically exfoliated into flakes of ~100 nm in thickness, Pb thin films (~200 nm thick) and/or Pb grains (10-50 nm in diameter and 10-20 nm in height) were deposited to the surface of the flakes via electron beam lithography and sputtering procedures. The Pb grains were formed naturally on the $Bi_2Te_3$ surface by controlling the deposition time. Figure S1 shows the optical microscope images of typical Pb-grain devices.

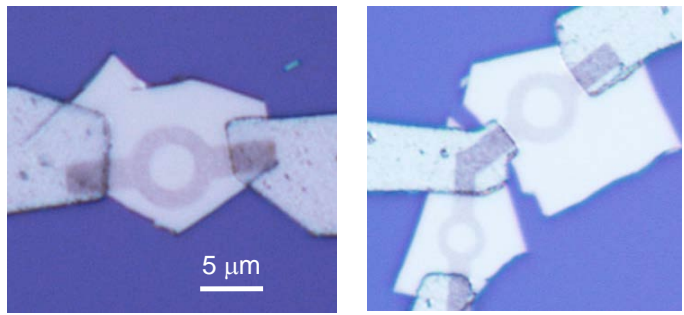

**Figure S1 |** Optical microscope images of typical Pb-grain SQUIDs.

2. More devices with different Pb coverage ratios

The central idea of constructing an arbitrary-phase loop using superconducting proximity effect is to cut the *s*-wave superconductor into small segments and rearrange them on the TI surface to form an interference loop. In order to compromise between the strength of proximity effect and the amount of phase shift along the loop, we have tried several different Pb coverage ratios by varying the Pb sputtering time. The coverage ratio can be determined from the SEM pictures of the devices by using software such as Photoshop.

It has to be noted that the sputtering time window for controlling the coverage ratio of Pb in our sputtering machine is only ~ 10 seconds, and that there is not a linear relationship between the deposition time (hence the nominal thickness of the Pb film) and the coverage ratio yielded – it seems that the transition from forming discrete Pb grains to forming a connected network takes place rather abruptly. Adding to the difficulties, the morphology of the Pb grains might also depend on the surface cleanness of the $Bi_2Te_3$ flakes, the vacuum while depositing Pb, etc., although we tried to maintain all the conditions the same throughout the experiment. Nevertheless, it would still be helpful to show the statistical data of our investigation.

Overall we have investigated more than ten Pb-grain devices in four batches with different Pb coverage ratios. Seven of them showed pattern B. The results are summarized in Table S1.

Table S1 | Statistics on devices with different Pb coverage ratio.

| Sputtering Time | >90 s | 10.5-11 s | 9.5-10 s | 6.5-7 s ** | 9-9.5 s | 8-8.5 s |
|---|---|---|---|---|---|---|
| Pb Coverage Ratios | Standard SQUID | >90% | 71±2% | 74±2% | 62±3% | 58±3% |
| # of Devices | > 15 | 3 | 1 | 1 | 3 | 2 |
| SEM Pictures | Continues Pb Film | 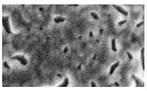 | 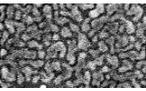 | 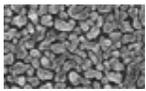 | 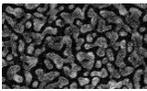 | 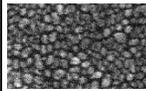 |
| Pattern A | Yes $I_c$ =1- 10 μA | Yes $I_c$ =0.3- 0.7 μA * | Yes $I_c$ ~ 0.5 μA * | Yes $I_c$ ~ 0.5 μA * | No | No |
| Pattern B | No | No | Yes $I_c$ ~ 0.5 μA * $\delta$ ~ 0.37π-0.60π | Yes $I_c$ ~ 0.75 μA * $\delta$ ~ 0.87π | Yes $I_c$ ~ 1.0 μA * $\delta$ ~ 0.69π | Yes $I_c$ ~ 0.1 μA * |

\* The characteristic currents depend on the detailed geometry of the devices.
\*\* This batch of devices were fabricated with slightly varied microfabrication procedures.

It turns out that once the Pb grains are connected to form a network, which usually happens when the sputtering time is longer than ~11 s on a sputtering rate of ~1.7 nm/s in our experiment, the interference pattern of the device becomes the same as that of the continuous Pb film-$Bi_2Te_3$ SQUIDs reported in Ref. [1]. No pattern B was observed in this situation even if the pattern A of the devices was suppressed in a parallel magnetic field of > 300 G, as shown in Fig. S2.

Shortening the sputtering time/lowering the Pb coverage ratio reduced the amplitude of pattern A to be below 1 μA, and allowed pattern B to emerge. With further lowering the Pb coverage ratio, only pattern B remained, as summarized in Table S1.

So far we are unable to control the phase shift of pattern B from device to device systematically.

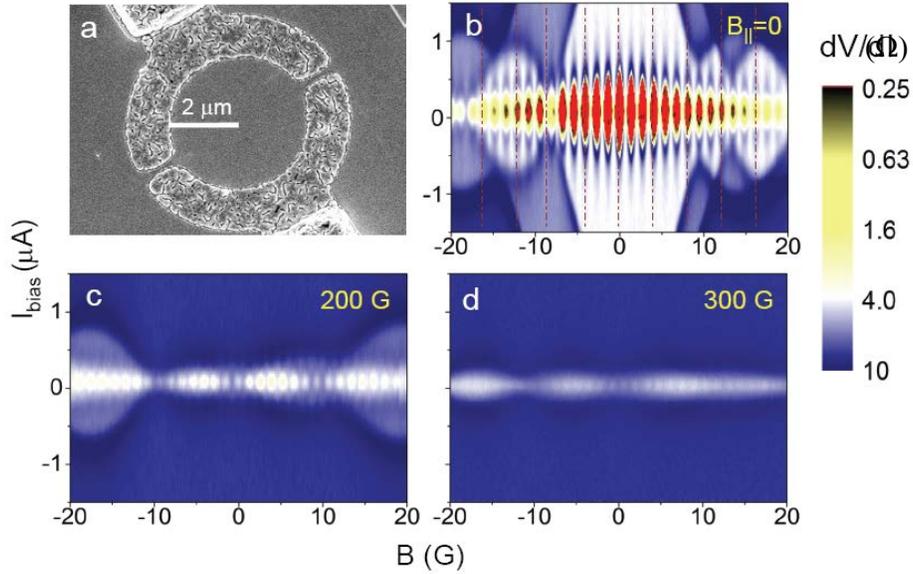

**Figure S2** | Pattern A of a Pb network SQUID (Pb coverage ratio >90%) measured at 30 mK and in parallel magnetic fields. The zero resistance state is marked in red color (dV/dI<0.2 Ω). The interference pattern was regular and unshifted/untilted. The pattern was suppressed in a parallel magnetic field of > 300 G. No pattern B was observed.

Figures S2-1, S2-2 and S2-3 show the results obtained on another device which has a geometry similar to the one shown in Fig. 2a of the main manuscript. It reproduced the situation that a normal pattern A and an anomalous pattern B coexist. Since these results were reproduced by a different person, the dependences between Pb sputtering time and grain thickness and coverage ratio were also different, although the sputtering rate was kept the same as before.

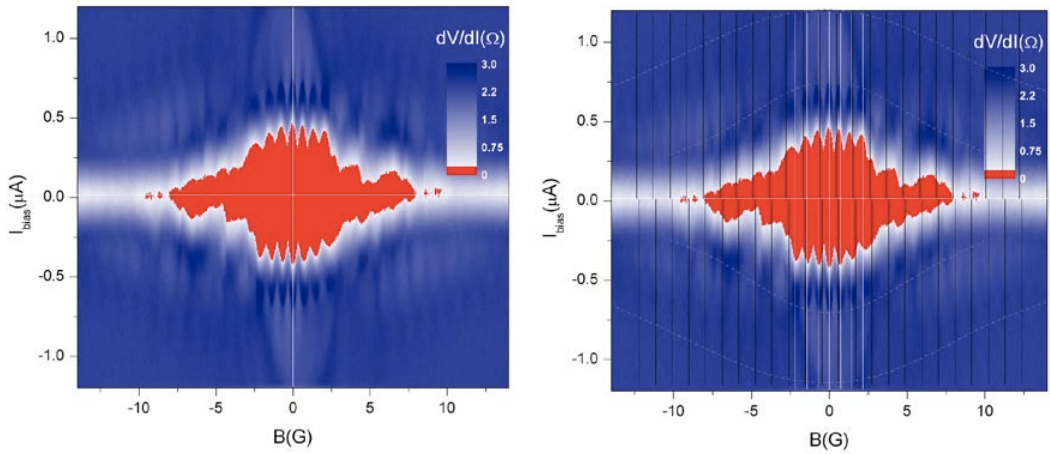

**Figure S2-1** | Patterns A and B observed on another Pb-grain SQUID measured at 30 mK. The zero resistance state is marked in red color (dV/dI<0.2 Ω). Left: original data. Right: with lines guiding to the eyes. The dashed lines mark the envelops of two independent interference patterns. The vertical white/black lines illustrate the peak position of pattern A/B. The peak positions in pattern B are horizontally shifted by $\delta \approx 0.87\pi$ between its positive and negative bias current directions. And the period of pattern B is ~40% larger than that of pattern A.

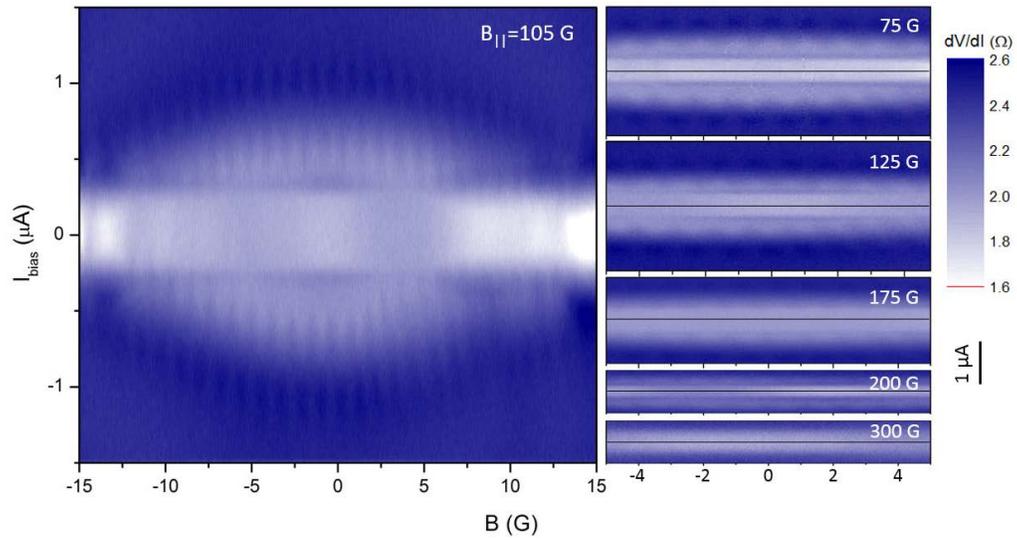

**Figure S2-2 |** Parallel magnetic field dependence of the patterns shown in Fig. S2-1. T=30 mK. Pattern A was entirely suppressed in a parallel magnetic field of ~75 G, so that a uniform pattern B emerged (left panel). Pattern B survives in parallel magnetic fields up to ~175 G (right panel, where the vertical scale is separately indicated by the black bar).

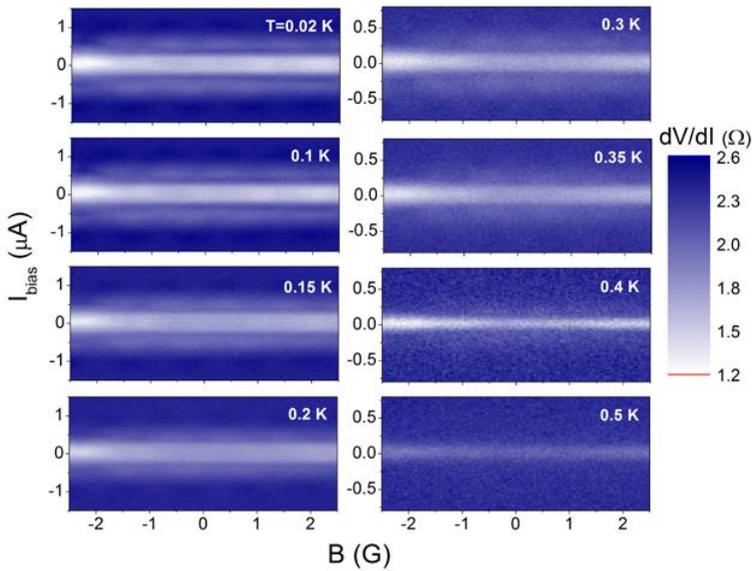

**Figure S2-3 |** Temperature dependence of the amplitude of pattern B in a parallel magnetic field of ~75 G (upper panel), with a line shape indicating that the Josephson junctions are in a state close to the dirty limit (lower panel).

3. The upper critical fields of Pb thin films and Pb grains

In the main frame of Fig. S3 we show the resistance of a typical Pb-grain device (the second device shown in Fig. 4 of the main manuscript) as a function of magnetic field applied parallel to the surface of the $Bi_2Te_3$ flake. One can see that the total resistance goes up when the parallel magnetic field excesses the upper critical field of the Pb film, ~2000 G, as marked by black arrows. And there are another upturns starting at $\pm1.4$ T, indicated by the red arrows, corresponding to the lower bound of the upper critical fields of the Pb grains. The inset of Fig. S3 shows the resistance transition of a Pb electrode (a continuous Pb film) of that device, whose critical temperature is 7.2 K (not shown), and the upper critical field is about 0.2 T at 30 mK, as indicated by the black arrows.

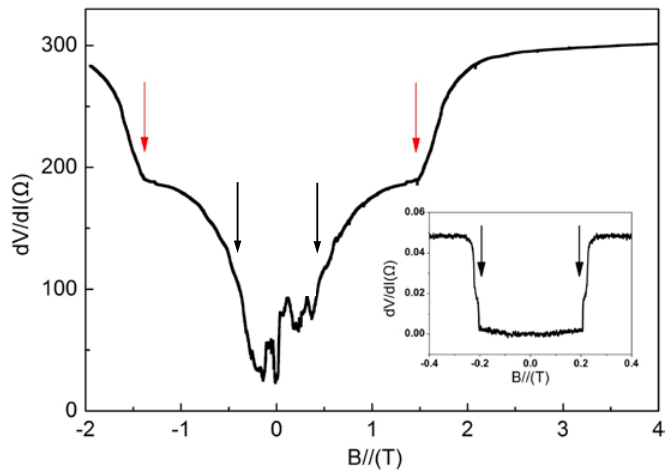

**Figure S3 |** (main frame) The parallel magnetic field dependence of the resistance of the second device presented in the main manuscript, measured at 30 mK. Inset, the parallel magnetic field dependence of the resistance of a continuous Pb electrode of that device, measured in a 3-probe configuration at 30 mK.

4. Interplay and competition between pattern A and pattern B

When bulk superconductivity dominates along the ring, pattern A is the only interference pattern. When the bulk superconductivity subsides in-between the Pb islands, the effect of surface superconductivity emerges, thus anomalous interference of Cooper pairs occurs.

The amplitude of pattern B in this experiment is in sub-μA range. This pattern is observable only when pattern A's amplitude is reduced to the same range or even smaller. When the amplitudes of the two patterns are comparable, further suppression of the superconducting bulk by applying a parallel magnetic field would change the overall ratio of the s-wave segment to the p-wave-like segment. As a result, the relative shift between the positive and negative current halves would vary within a certain range of parallel magnetic field. Figure S4 shows the evolution of the phase shift with parallel magnetic field for the first device presented in the main manuscript.

When the amplitude of pattern A is reduced to be comparable with that of the pattern B, its shape is influenced by the latter, so that the zero-resistance state takes a zigzag shape as shown in Figs. 3b and 3h of the main manuscript.

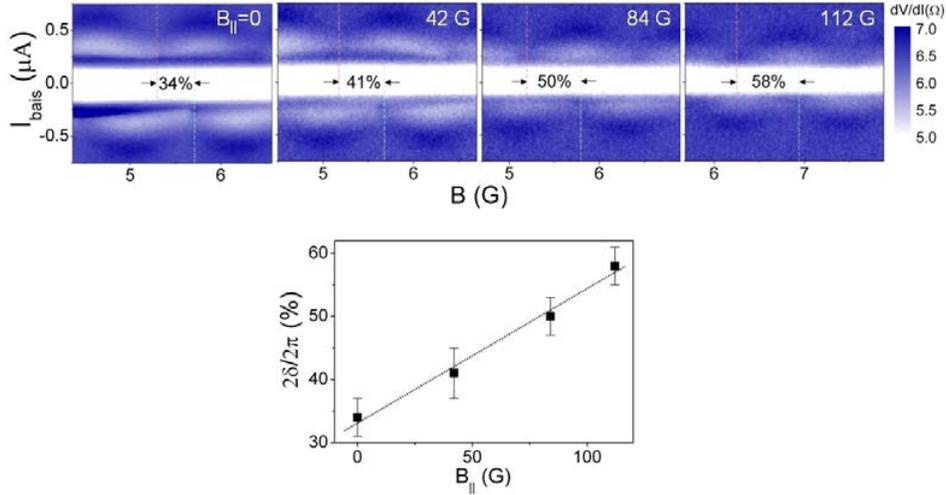

**Figure S4** | (Upper panel) The relative shift between the positive and negative current halves of pattern B as a function of parallel magnetic field marked in the frames, observed at 30 mK on the first device in the main manuscript. The lines and the numbers denote the phase shift $2\delta$ as a percentage of $2\pi$. (Lower panel) A summary of the phase shift percentages as a function of parallel magnetic field.


5. Investigation on a four-segment SQUID

Shown in Fig. S5 is the Cooper pair interference pattern measured from the two opposite electrodes of a four-segment SQUID, following the design illustrated in Fig. 1d. Similar to Pb-grain SQUIDs, two distinctive patterns were observed. Pattern B, the one with whitish-blue color inhabited on a finite resistance state, is oppositely shifted between its upper and lower halves, regardless that this device is geometrically symmetric and showing an unshifted pattern A.

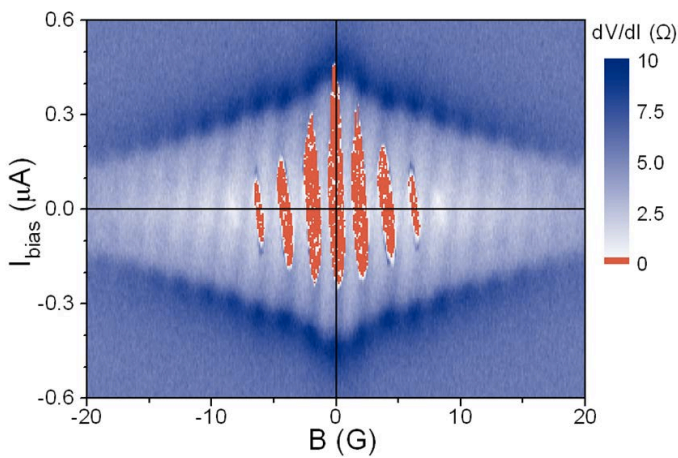

**Figure S5** | Differential resistance of a symmetric four-segment SQUID (following the design illustrated in Fig. 1**d**) measured at 30 mK, as functions of bias current $I_{bias}$ and magnetic field B perpendicular to the device plane. Similar to the data of Pb-grain devices, two oscillation patterns can be recognized. Pattern A (colored in red) is mostly unshifted/untilted. Its asymmetry in the vertical direction is due to the existence of hysteresis in V-$I_{bias}$ curve. Pattern B (in whitish-blue color) appears to be shifted/tilted.

6. Is pattern B shifted or tilted?

At a first glance, pattern B in Fig. 4 of the main manuscript and Fig. S5 above appears to be tilted. However, this is an illusion. Since the peak height of pattern B in these figures varies significantly from the center to the side, it allows us to distinguish whether the pattern is tilted, or rather being shifted between the positive and negative current halves. Taking the data in Fig. S5 as an example and replotted them in Fig. S6a, if the interference patterns are tilted in same slope, then the peak positions should follow the tilted parallel lines. This is however not true. Obvious deviations can be found at the positions pointed by the arrows. In fact, the peak positions keep uniformly distributed regardless of the height of each peak, as illustrated with the help of two sets (upper and lower) of evenly spaced vertical lines in Fig. S6b.

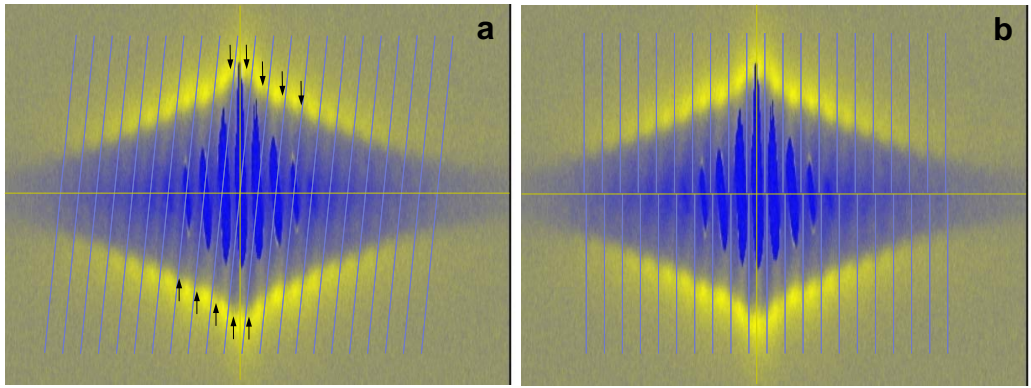

**Figure S6 |** The interference patterns of a symmetric four-junction SQUID shown in Fig. S5. The assistive lines help to show that the pattern B is shifted between its positive and negative current directions, rather than being tilted. The arrows in the left panel indicate that the peak positions deviate from the tilting trend.

7. Spontaneous symmetry breaking and the current-driven symmetry breaking in a double-well system

After added a phase shift of $\pm\delta$ to the phase quantization condition, the free energy minimum of a SQUID shifts from zero to finite positive/negative magnetic fields, corresponding to clockwise/counterclockwise (cw/ccw) circulation modes. These two modes are degenerate in energy, forming a quantum double-well as illustrated in Fig. S7.

If there is no inter-well tunneling, then the system undergoes spontaneous symmetry breaking by picking up one of the well states, accompanied with a chiral edge current, just like in

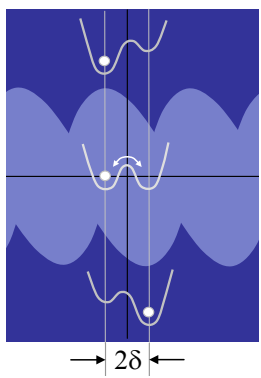

**Figure S7 |** A double-well potential for the clockwise and counterclockwise circulation modes of an arbitrary-phase loop. The horizontal and vertical axes are magnetic flux and bias current, respectively. Time reversal symmetry is kept when a two-level system is formed due to inter-well tunneling.

a π-loop ring made of time-reversal invariant s- and d-wave superconductors. In this case, the pattern is shifted by a same amount either to the positive or to the negative field direction, in both the positive and negative bias current directions if the bias current is not too large, so that the pattern keeps untilted. One should be able to distinguish the shifting, if the position of zero magnetic field is accurately known to locate not at the maximum of the patterns.

If the tunneling between the cw and ccw double wells is significant, which usually happens when the inter-well barrier is low, such that macroscopic quantum tunneling between the wells occurs, the device further forms a two-level system (TLS), in which both the ground and first excitation levels are linear combinations of the cw and ccw modes. In this case, the time-reversal symmetry is restored such that $I_c(-B)=-I_c(B)$.

Applying a bias current not only tilts the washboard potential along the $\phi_1-\phi_2$ direction in the parametric space, but also introduces asymmetry to the double-well potential along the $\phi_1+\phi_2$ direction via a given asymmetric current distribution of the device (where $\phi_1$ and $\phi_2$ are the phase differences across junctions 1 and 2 defined along the circulation direction). It breaks the symmetry between the two wells, so that the device tends to dwell longer in one of the wells selected by the direction of the bias current, leading to the observed horizontal shift of pattern B between its positive and negative bias current directions.

The 2D energy profile and its tilting in the presence of a bias current is illustrated in Fig. S7-1.

The finite tunneling rate along the washboard potential direction gives rise to the dissipation and a finite resistance in the pattern B region. The voltage can be estimated as: V = $(h/2e)<d(\phi_1-\phi_2)/dt>$. According to the data in Figs. 2 d and e, and those in Figs. S2-1 and S5, V is a few μV, which is comparable to the thermal energy of ~30 mK in the measurement, indicating probably a thermally-assisted phase diffusion mechanism for the finite resistance state of pattern B.

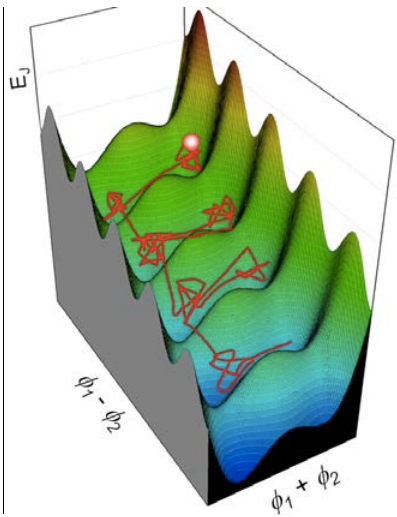

**Figure S7-1 |** A bias current tilts the 2D energy profile not only along the washboard potential ($\phi_1-\phi_2$) direction, but possibly also along the cw and ccw double-well ($\phi_1+\phi_2$) direction (depending on if the device is geometrically asymmetric). Phase diffusion along the former direction is dissipative. The red path illustrates the macroscopic tunneling and trapping processes of the phase particle.

The phase changing rate $<d(\phi_1-\phi_2)/dt>$, with an effective rate of ~ 1 GHz, is not the intrinsic free-running frequency, but an average over running and trapping. The trapping/dwelling time within one well is roughly 1 nS, a time long enough for the Cooper pair to establish interference along the mesoscopic-sized ring, so that pattern B is still observable.

8. Further explanations on the Berry phase and the possible consequences of having arbitrary phase loops in mesoscopic hybrid rings

Winding the mass center of a Cooper pair rotates the two spins of paired helical electrons in the same angular direction (Figs. S8a and S8b), thereby doubles the total Berry phase accumulated. This is true no matter whether the total spin is S=0 (singlet pairing) or S=1 (triplet pairing), nor does it matter whether the Cooper pair is wound adiabatically slow or fast compared to the Fermi velocity $v_F$. The reason that the total Berry phase is doubled instead of being cancelled in the S=0 case is that, due to the Berry curvature, the two anti-parallel spins with opposite $v_F$ but same helicity feel opposite effective magnetic fields individually, rather than a same external magnetic field.

In an alternative language, the Copper pair "lives" in a curved one-dimensional space on the ring. It experiences a geometric phase if the electron spins prefer the radiant direction.

A Berry phase of $2\pi$ over 360° on a 2D surface gives rise to nothing but the $p_x+ip_y$ symmetry.

As mentioned before, if there is tunneling between the cw and ccw modes, then the hybrid ring is in a state which is the linear combination of the two modes, so that the TRS is kept. We believe that our experimental case is in this regime. However, if there is no inter-well tunneling, then the TRS will be broken spontaneously, and the segment on TI surface will behave like a true $p_x+ip_y$ superconductor. The following discussions refer mostly to this regime.

The Cooper pair interference devices constructed in the manner proposed in this work are generally arbitrary phase loops, because the total Berry phase encountered by the Cooper pair over one turn along the ring depends on the turning angle of the mode in the $p_x+ip_y$ segment. For arbitrary-phase loops with irrational ratio of the two segments in general, mode-locking to its nearest rational number would occur in the presence of non-linearity caused by, for example, electron-electron interaction. We note that mode-locking is a common phenomenon in non-linear systems.

As a mesoscopic hybrid system, an arbitrary phase loop could be thought of as an artificial molecule made of different artificial atoms (segments), and having its own unique ground state(s).

On a mesoscopic $\pi$–loop ring, i.e., half made of s-wave superconductor and half made of $p_x+ip_y$-wave superconductor, as shown in Figs. 1b and S8a, a Cooper pair needs to wind twice to pick up a total Berry phase of $2\pi$ in the non-tunneling limit (between the two wells), in order to satisfy the phase quantization condition (Fig. S8c, where we assume that the spins do not rotate in the conventional s-wave segment where the Berry curvature is zero). Such mode has a winding number (topological quantum number) of 2, referred to as the 1/2 fractional quantum mode. Similarly, a loop whose 1/3 is made of $p_x+ip_y$-wave superconductor would host the 1/3 fractional quantum mode, with a winding number of 3, i.e., the Cooper pair needs to wind three turns to pick up a total Berry phase of $2\pi$ in order to satisfy the phase quantization condition (Fig. S8d). In general, a mesoscopic ring whose 1/n is made of $p_x+ip_y$ segment requires the Cooper pair to wind n turns in order to form the $n^{th}$ fractional quantum mode.

We call them fractional quantum modes because the wavelength of the modes is n times longer than that of n=1 mode, in contrast to high angular momentum modes whose wavelength is 1/n of the n=1 mode.

Once the concept of fractional quantum mode is established, one might have to deal with the

concept of fractional Cooper pairs on mesoscopic hybrid rings. While a rigorous exploration should be the task of theorists, in the following we would try to present a qualitative description from an experimentalist's point of view.

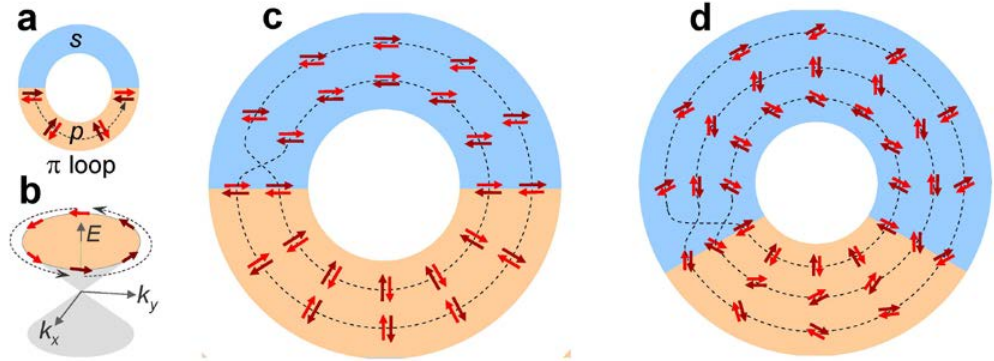

**Figure S8 | a,** Illustration of a π-ring. The light-blue segment is made of conventional *s*-wave superconductor, and the orange segment is made of $p_x+ip_y$-wave superconductor. **b,** Winding the mass center of a Cooper pair over the $p_x+ip_y$-wave half ring rotates the two spins of paired helical electrons along the same angular direction, picking up a Berry phase of π. **c,** The 1/2 fractional quantum mode hosted on the π-ring, with a winding number of 2, i.e., the Cooper pair there needs to wind twice as if on a Mobius strip, to complete the formation of the mode. **d,** The 1/3 fractional quantum mode with a winding number of 3.

If neglecting the multi-segment details and attributing the Berry phase of (2π)/n acquired by 2e on the $p_x+ip_y$-like segment to the entire ring, then the effective charge of the Cooper pair appears to be (2e)/n. This assignment is made in comparison to the facts that a Berry phase of π is acquired by a single electron of charge e along a TI ring [2], and a Berry phase of 2π is acquired by a Cooper pair of charge 2e along a $p_x+ip_y$ superconducting ring. The fractionalization of Berry phase in the modes leads accordingly to the fractionalization of the charges.

To help understanding the fractionalization of Cooper pair, a 2e Cooper pair could be thought of as being fractionalized to n parts in the 1/n fractional quantum mode, with each part carrying a same fractional charge of (2e)/n but differing in phase by (2π)/n from each other, circulating in parallel along the ring as shown in Fig. S8c (n=2) and Fig. S8d (n=3).

A fractional Cooper pair with a charge of (2e)/n winding n turns to form a complete mode is in analogy with the anyons in a two-dimensional electron gas (2DEG). For example, a Laughlin anyon with a fractional charge of e/3 winds three turns to bind with a fluxoid, thus to form a complete mode there.

We note that the concept of fractional Cooper pair has previously been discussed in uniform systems such as 2DEGs [3-7] and TI [8]. Its generalization to mesoscopic hybrid systems, as discussed above, would warrant further studies both theoretically and experimentally.



\end{document}